\documentclass[a4paper]{jpconf}
\usepackage{ae}
\usepackage{bm}
\usepackage{color}
\usepackage{amssymb}
\usepackage{amsmath}
\usepackage{amsfonts}
\usepackage{graphicx}
\usepackage{slashed}
\usepackage{wasysym}
\usepackage{setspace}
\usepackage{subfigure}
\usepackage{ulem}
\usepackage{cite}

\newcommand{\lambdabar}{{\mkern0.75mu\mathchar '26\mkern -9.75mu\lambda}}

\newcommand{\Eqref}[1]{Eq.~\eqref{#1}}

\begin{document}

\title{Photon polarization tensor in circularly polarized Hermite- and Laguerre-Gaussian beams}

\author{Felix Karbstein$^{1,2}$ and Elena A. Mosman$^{3,4}$}

\address{$^1$ Helmholtz-Institut Jena, Fr\"obelstieg 3, 07743 Jena, Germany} 

\address{$^2$ Theoretisch-Physikalisches Institut, Abbe Center of Photonics, Friedrich-Schiller-Universit\"at Jena, Max-Wien-Platz 1, 07743 Jena, Germany}

\address{$^3$ National Research Tomsk Polytechnic University, Lenin Ave. 30, 634050 Tomsk, Russia}

\address{$^4$ Physics Faculty, Tomsk State University, Lenin Ave. 36, 634050 Tomsk, Russia}

\ead{felix.karbstein@uni-jena.de, mosmanea@tpu.ru}

\begin{abstract}
 We derive analytical expressions for the photon polarization tensor in circularly polarized Hermite- and Laguerre-Gaussian beams, complementing the corresponding results for linearly polarized beams obtained recently.
 As they are based upon a locally constant field approximation of the one-loop Heisenberg-Euler effective Lagrangian for quantum electrodynamics (QED) in constant fields, our results are generically limited to slowly varying electromagnetic fields, varying on spatial (temporal) scales much larger than the Compton wavelength (time) of the electron. 
\end{abstract}

\section{Introduction}

The photon polarization tensor is a central object in the study of probe photon propagation in the quantum vacuum.
It encodes quantum corrections to the classical propagation of light.
In the absence of electromagnetic fields, by construction quantum fluctuations do not change the vacuum speed of light and do not result in any modifications for photon propagation.
This is different in external fields: As electromagnetic fields couple to charges, they can {\it polarize} virtual charged particle-antiparticle fluctuations \cite{Heisenberg:1935qt,Weisskopf}, giving rise to remarkable effects, such as vacuum birefringence \cite{Toll:1952,Baier,BialynickaBirula:1970vy,Heinzl:2006xc,DiPiazza:2006pr,Dinu:2013gaa,Karbstein:2015xra,Schlenvoigt:2016,Karbstein:2016lby,Kotkin:1996nf,Nakamiya:2015pde,Ilderton:2016khs,King:2016jnl,Bragin:2017yau} and -- in inhomogeneous fields -- vacuum diffraction effects \cite{Hatsagortsyan:2011,King:2013am,Tommasini:2010fb,King:2012aw,Gies:2013yxa} experienced by probe photons; cf. also the reviews \cite{Dittrich:2000zu,Marklund:2008gj,DiPiazza:2011tq,Battesti:2012hf,King:2015tba} and references therein. 
For ongoing experimental efforts aiming at the detection of these effects, see Refs.~\cite{Cantatore:2008zz,Berceau:2011zz,Fan:2017fnd,Inada:2017lop}.
Exact analytical results for the photon polarization tensor $\Pi^{\rho\sigma}$ in external field QED
at one-loop level $\sim\alpha/\pi$, where\footnote{Throughout this article, we use the Heaviside-Lorentz system and units where $c=\hbar=1$. Our metric convention is $g^{\mu\nu}={\rm diag}(-1,+1,+1,+1)$.} $\alpha=e^2/(4\pi)\simeq1/137$ is the fine-structure constant, with elementary charge $e$, are available for constant electromagnetic fields \cite{BatShab,narozhnyi:1968} and plane wave backgrounds \cite{Baier:1975ff,Becker:1974en}.

Besides, approximate results for $\Pi^{\rho\sigma}$ in slowly varying inhomogeneous electromagnetic fields have been determined on the basis of a locally constant field approximation (LCFA) \cite{Karbstein:2015cpa} to the Heisenberg-Euler effective Lagrangian in constant fields \cite{Heisenberg:1935qt}.
For insights into the Heisenberg-Euler effective Lagrangian beyond constant fields, cf. the derivative expansion results \cite{Gusynin:1995bc,Dunne:2000up}.
The LCFA allows for reliable insights in electromagnetic field configurations which vary on spatial (temporal) scales much larger than the Compton wavelength $\lambdabar=1/m_e\approx3.86\cdot10^{-13}\,{\rm m}$ (time $\lambdabar/c\approx1.29\cdot10^{-21}\,{\rm s}$) of the electron; $m_e\approx511\,{\rm keV}$ is the electron mass.
So far, this approach has been adopted to determine $\Pi^{\rho\sigma}$ for linearly polarized paraxial Hermite- and Laguerre-Gaussian beams of arbitrary mode composition \cite{Karbstein:2017jgh}.

In these proceedings, we provide the analogous expressions for circular polarized beams.

\section{Results}

\subsection{Field configuration}

The electric and magnetic field vectors describing a generic superposition of co-propagating, electromagnetic waves in vacuum can be represented as 
\begin{equation}
 \vec{E}=\sum_{\cal N}{\cal E}_{\cal N}\bigl(\hat{\vec{e}}_{\rm x}\cos\phi_{\cal N}+\hat{\vec{e}}_{\rm y}\sin\phi_{{\cal N}}\bigr)\,,\quad
 \vec{B}=\vec{E}\big|_{\phi_{\cal N}\to\phi_{\cal N}+\frac{\pi}{2}}\,, \label{eq:Fields}
\end{equation}
where ${\cal E}_{\cal N}$ denote amplitude profile and $\phi_{\cal N}$ a phase function of mode $\cal N$, respectively.
Without loss of generality, in \Eqref{eq:Fields} we assumed the waves to propagate along the positive $\rm z$ axis.
Aiming at the determination of $\Pi^{\rho\sigma}$ in this field configuration, it is helpful to note that \Eqref{eq:Fields} implies
\begin{align}
    {\cal F}&=\frac{1}{4}F_{\mu\nu}F^{\mu\nu}
    =0\,,\nonumber \\
    {\cal G}&=\frac{1}{4}F_{\mu\nu}{}^*\!F^{\mu\nu}=-\sum_{\cal N}\sum_{{\cal N}'}{\cal E}_{\cal N}{\cal E}_{{\cal N}'}\sin(\phi_{\cal N}-\phi_{{\cal N}'}) , \nonumber\\
    (k F)^\rho&=\sum_{\cal N}{\cal E}_{\cal N}\bigl[\varepsilon_2^\rho(k)\sin\phi_{\cal N}+\varepsilon_1^\rho(k)\cos\phi_{\cal N}\bigr], \nonumber \\
    (k{}^*\!F)^\rho&=\sum_{\cal N}{\cal E}_{\cal N}\bigl[\varepsilon_2^\rho(k)\cos\phi_{\cal N}-\varepsilon_1^\rho(k)\sin\phi_{\cal N}\bigr], \label{eq:bb}
\end{align}
with $k^\mu=(\omega,\vec{k})$, $(kF)^\mu=k_\nu F^{\nu\mu}$
and dual field strength tensor ${}^*\!F^{\mu\nu}=\frac{1}{2}\epsilon^{\mu\nu\alpha\beta}F_{\alpha\beta}$.
The four-vectors $\varepsilon_{1,2}^\mu(k)$ are defined as \cite{Karbstein:2015cpa}
\begin{equation}
 \varepsilon_1^\mu(k):=(-k_{\rm x},k_{\rm z}-\omega,0,-k_{\rm x})
 \,,\quad
 \varepsilon_2^\mu(k):=(-k_{\rm y},0,k_{\rm z}-\omega,-k_{\rm y})\,.
\end{equation}

\subsection{Photon polarization tensor}

Resorting to a LCFA, the one-loop photon polarization tensor in the background field~\eqref{eq:Fields} can be expressed as \cite{Karbstein:2015cpa}\footnote{For completeness, note that this definition of $\Pi^{\rho\sigma}$ differs from that of Ref.~\cite{Karbstein:2015cpa} by an overall minus sign.}
\begin{equation}
 \Pi^{\rho\sigma}(k,k')
 =-\frac{\alpha}{\pi}\frac{1}{45}\Bigl(\frac{e}{m_e^2}\Bigr)^2\int{\rm d}^4x\,{\rm e}^{{\rm i}(k+k')x} \Bigl[4 (kF)^\rho  (k'F)^\sigma
 + 7 (k{}^*\!F)^\rho (k'{}^*\!F)^\sigma+7{\cal G}\,
 k'_\mu k_\alpha \epsilon^{\rho\sigma\mu\alpha}
 \Bigr], \label{eq:Pi}
\end{equation}
where we neglected terms $\sim\upsilon^2\bigl[{\cal O}\bigl((\frac{\upsilon}{m_e})^2\bigr)+{\cal O}\bigl((\frac{eF}{m_e^2})^4\bigr)\bigr]$. Here, the momentum scale $\upsilon$ delimits the typical momentum scales of variation of the background fields as well as the photon energies $\omega,\omega'$ and momenta $|\vec{k}|,|\vec{k}'|$ from above \cite{Galtsov:1982,Karbstein:2015cpa}.
With the help of \Eqref{eq:bb}, \Eqref{eq:Pi} can be cast into the following form
\begin{align}
 \Pi^{\rho\sigma}(k,k') = -\frac{\alpha}{\pi}\frac{1}{45}\sum_{\ell=\pm1}
  \biggl\{&\!-\frac{7}{2}\,\Pi^-_\ell(k+k') (-{\rm i})^l k'_\mu k_\alpha\epsilon^{\rho\sigma\mu\alpha} \nonumber\\
  +\frac{11}{4}\,\Pi_\ell^-(k+k')&\Bigl[\varepsilon_1^\rho(k)\varepsilon_1^\sigma(k')+\varepsilon_2^\rho(k)\varepsilon_2^\sigma(k')
  +(-{\rm i})^l\bigl[\varepsilon_2^\rho(k)\varepsilon_1^\sigma(k')
 -\varepsilon_1^\rho(k)\varepsilon_2^\sigma(k')]\Bigr] \nonumber\\
  -\frac{3}{4}\,\Pi^+_\ell(k+k')&\Bigl[\varepsilon_1^\rho(k)\varepsilon_1^\sigma(k')-\varepsilon_2^\rho(k)\varepsilon_2^\sigma(k')
  +(-{\rm i})^l\bigl[\varepsilon_2^\rho(k)\varepsilon_1^\sigma(k')+\varepsilon_1^\rho(k)\varepsilon_2^\sigma(k')\bigr]\Bigr] \biggr\} , \label{eq:Pi2}
\end{align}
where we made use of the definitions
\begin{equation}
 \Pi_\ell^\pm(k+k') = \sum_{\cal N}\sum_{{\cal N}'}\,(\Pi_\ell^\pm)_{{\cal N};{\cal N}'}(k+k')\,, \label{eq:Pi2b}
\end{equation}
and
\begin{equation}
 (\Pi_\ell^\pm)_{{\cal N};{\cal N}'}(k+k')=\int{\rm d}^4x\,{\rm e}^{{\rm i}(k+k')x}\,\frac{e{\cal E}_{\cal N}}{m_e^2}\,\frac{e{\cal E}_{{\cal N}'}}{m_e^2}\,{\rm e}^{{\rm i}\ell(\phi_{\cal N}\pm\phi_{{\cal N}'})} \,. \label{eq:Pi2c}
\end{equation}

\subsection{Circular polarization}

The above results can be straightforwardly adopted to describe the photon polarization tensor in circularly polarized paraxial beams at leading order in the diffraction angle $\theta\ll1$ (cf. below) \cite{Davis:1979,Barton:1989,Salamin:2006ff}.
In this case, we have
\begin{align}
 {\cal E}_{\cal N}\quad&\to\quad{\cal E}_{\cal N}(x)\,, \nonumber\\
 \phi_{\cal N}\quad&\to\quad\pm\Phi_{\cal N}(x)\,,
\end{align}
where $+/-$ refers to right-/left-handed circular polarization. For the explicit mode amplitude ${\cal E}_{\cal N}(x)$ and phase profiles $\phi_{\cal N}(x)$ adopted in these proceedings, see below. 
Hence, in our explicit calculations we can without loss of generality stick to right-handed circular polarization, $\phi_{\cal N}\to\Phi_{\cal N}(x)$.
The result for left-handed circular polarization follows therefrom by substituting $\Pi_\ell^\pm(k+k')\to\Pi_{-\ell}^\pm(k+k')$ in \Eqref{eq:Pi2}.

\subsection{Field profiles}

Hermite-Gaussian (HG) modes are labeled by ${\cal N}=\{m\in\mathbb{N}_0,n\in\mathbb{N}_0\}$. They exhibit a Cartesian symmetry about the beam's propagation direction and are characterized by the following amplitude and phase profiles \cite{Siegman},
\begin{align}
 {\cal E}_{m,n}(x)&={\mathfrak E}_{m,n}\,H_m\Bigl(\tfrac{\sqrt{2}{\rm x}}{w({\rm z})}\Bigr)H_n\Bigl(\tfrac{\sqrt{2}{\rm y}}{w({\rm z})}\Bigr)\,{\rm e}^{-\frac{({\rm z}-t)^2}{(\tau/2)^2}}\,\frac{w_0}{w({\rm z})}\,{\rm e}^{-\frac{r^2}{w^2({\rm z})}}, \nonumber\\
  \Phi_{m,n}(x)&=
    \Omega({\rm z}-t)+\frac{\rm z}{{\rm z}_R}\frac{r^2}{w^2({\rm z})}-(m+n+1)\arctan\Bigl(\frac{\rm z}{{\rm z}_R}\Bigr)+\varphi_{n,m}\,, \label{eq:HG}
\end{align}
where $H_m(\chi)$ denote Hermite polynomials, $\Omega=2\pi/\lambda$ is the frequency of the beam, and $w({\rm z})=w_0\sqrt{1+({\rm z}/{\rm z}_R)^2}$, with waist size $w_0$ and Rayleigh range ${\rm z}_R=\pi w_0^2/\lambda$.
The constants ${\mathfrak E}_{\cal N}$ and $\varphi_{\cal N}$ are a mode specific peak field amplitude and phase offset, respectively, and $\tau$ is the pulse duration.

Analogously, Laguerre-Gaussian (LG) modes are labeled by ${\cal N}=\{l\in\mathbb{Z},p\in\mathbb{N}_0\}$. The transverse profiles of LG modes are circularly symmetric and, resorting to polar coordinates, $r=\sqrt{{\rm x}^2+{\rm y}^2}$ and $\varphi={\rm arg}({\rm x}+{\rm iy})$, read  \cite{Siegman}
\begin{align}
 {\cal E}_{l,p}(x)&={\mathfrak E}_{l,p} \biggl(\frac{\sqrt{2}r}{w({\rm z})}\biggr)^{|l|} L^{|l|}_p\Bigl(\bigl(\tfrac{\sqrt{2}r}{w({\rm z})}\bigr)^2\Bigr) \,{\rm e}^{-\frac{({\rm z}-t)^2}{(\tau/2)^2}}\,\frac{w_0}{w({\rm z})}\,{\rm e}^{-\frac{r^2}{w^2({\rm z})}}, \nonumber\\
  \Phi_{l,p}(x)&=
    \Omega({\rm z}-t)+\frac{\rm z}{{\rm z}_R}\frac{r^2}{w^2({\rm z})}-(|l|+2p+1)\arctan\Bigl(\frac{\rm z}{{\rm z}_R}\Bigr)-l\varphi+\varphi_{l,p}\,, \label{eq:LG}
\end{align}
where $L_p^{|l|}(\chi)$ are generalized Laguerre polynomials. For further details, cf. Ref.~\cite{Karbstein:2017jgh}.

For given laser parameters, i.e., total pulse energy $W$, pulse duration $\tau$ and waist radius of the fundamental Gaussian mode $w_0$, the total energy is partitioned into the different modes ${\cal N}$ as $W=\sum_{\cal N}W_{\cal N}$, where $W_{\cal N}$ denotes the energy put in mode $\cal N$.
Making use of the fact that the intensity associated with the electromagnetic field in mode $\cal N$ is given by $I_{\cal N}=\vec{E}_{\cal N}^2={\cal E}_{\cal N}^2$, where $\vec{E}_{\cal N}={\cal E}_{\cal N}(\hat{\vec{e}}_{\rm x}\cos\phi_{\cal N}+\hat{\vec{e}}_{\rm y}\sin\phi_{{\cal N}})$, it is straightforward to derive the following relation between $W_{\cal N}$ and the peak field amplitude $\mathfrak{E}_{\cal N}$,
\begin{equation}
W_{\cal N}=\Bigl(\frac{\pi}{2}\Bigr)^{\frac{3}{2}}\,{\mathfrak c}_{\cal N}\,{\mathfrak E}_{\cal N}^2\,\frac{\tau}{2}\,w_0^2
\quad\leftrightarrow\quad
{\mathfrak E}_{\cal N}^2=4\sqrt{\frac{2}{\pi}}\frac{1}{{\mathfrak c}_{\cal N}}\frac{W_{\cal N}}{\pi w_0^2\tau}\,,
\label{eq:Wpulse}
\end{equation}
with mode specific coefficients ${\mathfrak c}_{\cal N}$ given by $c_{\cal N}\to{\mathfrak c}_{m,n}=2^{m+n}\,m!\,n!$ for HG modes,
and $c_{\cal N}\to{\mathfrak c}_{p,l}=(p+|l|)!/p!$ for LG modes.
For the details of the calculation, cf. the appendix of Ref.~\cite{Karbstein:2017jgh}.
We emphasize that the peak field amplitude squared in \Eqref{eq:Wpulse} amounts to half the one obtained for linearly polarized beams in Ref.~\cite{Karbstein:2017jgh}. The reason for this is that, in contrast to linear polarization, for circular polarization the phase factor $\phi_{\cal N}$ drops out completely when determining the mode intensity.

Inserting Eqs.~\eqref{eq:HG} and \eqref{eq:LG} into Eqs.~\eqref{eq:Pi2b} and \Eqref{eq:Pi2c} and performing the Fourier integrations, we obtain the following results for $\Pi^{\rho\sigma}$ in pulsed, circularly polarized LG and HG beams.

\subsection{Explicit results}

Following Ref.~\cite{Karbstein:2017jgh}, we represent the scalar coefficients~\eqref{eq:Pi2c} of the photon polarization as
\begin{align}
 (\Pi_\ell^\pm)_{{\cal N};{\cal N}'}(k+k')= \frac{e{\mathfrak E}_{\cal N}}{m_e^2}\frac{e{\mathfrak E}_{{\cal N}'}}{m_e^2}\,(2{\rm z}_R \, \pi w_0^2)\,
 \frac{\tau}{2}\sqrt{\frac{\pi}{2}}\,{\rm e}^{-\frac{1}{8}(\frac{\tau}{2})^2[\omega+\omega'+\ell(1\pm1)\Omega]^2}\,({\cal I}_\ell^\pm)_{{\cal N};{\cal N}'}(k+k') \,, \label{eq:Pi2c2}
\end{align}
where we encode the nontrivial momentum dependencies in the $\tau$ independent functions $({\cal I}_\ell^\pm)_{{\cal N};{\cal N}'}(k+k')$; note that $\lim_{\tau\to\infty}\frac{\tau}{2}\sqrt{\frac{\pi}{2}}\,{\rm e}^{-\frac{1}{8}(\frac{\tau}{2})^2\phi^2} =  2\pi\,\delta(\phi)$.
We will moreover make use of the definition
\begin{align}
 F_\Lambda\bigl(|a|,b\bigr):&=\int_{-\infty}^\infty\frac{\rm dz}{{\rm z}_R}\,
 \Bigl(\frac{w_0}{w({\rm z})}\Bigr)^\Lambda \,{\rm e}^{-{\rm i}a\frac{\rm z}{{\rm z}_R}-b\,(\frac{w({\rm z})}{w_0})^2} \nonumber\\
 &= \delta_{0,\Lambda}\sqrt{\frac{\pi}{b}}\,{\rm e}^{-\frac{1}{b}(\frac{a}{2})^2-b} + (1-\delta_{0,\Lambda}) \frac{\sqrt{\pi}}{\Gamma(\frac{\Lambda}{2})}\int_0^\infty\frac{{\rm d}s}{s}\,\frac{s^{\frac{\Lambda}{2}}}{\sqrt{s+b}}\,{\rm e}^{-\frac{1}{s+b}(\frac{a}{2})^2-(s+b)} \,, \label{eq:F}
\end{align}
for $b\geq0$, where $\delta_{0,\Lambda}$ is the Kronecker delta.
Equation~\eqref{eq:F} fulfills \cite{Karbstein:2017jgh},
\begin{equation}
 F_\Lambda\bigl(|a|,0\bigr)
 = \delta_{0,\Lambda}\, 2\pi\delta(a) + (1-\delta_{0,\Lambda})\, \frac{2\sqrt{\pi}}{\Gamma(\frac{\Lambda}{2})}\Bigl(\frac{|a|}{2}\Bigr)^{\frac{\Lambda-1}{2}}{\rm K}_{\frac{\Lambda-1}{2}}(|a|)\,,
 \label{eq:Fs}
\end{equation}
written in terms of the  Gamma function and the modified Bessel function of the second kind.
Employing the series representations of the Hermite polynomials and the generalized Laguerre polynomials as in Ref.~\cite{Karbstein:2017jgh}, the Fourier integrations can be performed explicitly and written in terms of $F_\Lambda\bigl(|a|,b\bigr)$ and parameter differentiations thereof.

For {\bf Hermite-Gaussian beams} we have ${\cal N}=\{m,n\}$, $N=m+n$, and obtain
\begin{align}
 ({\cal I}^\pm_\ell)_{{\cal N};{\cal N}'}^{{\rm HG}}(k)
 &=\frac{1}{4}\sum_{j=0}^{\lfloor\frac{m}{2}\rfloor}\sum_{q=0}^{\lfloor\frac{n}{2}\rfloor}\sum_{j'=0}^{\lfloor\frac{m'}{2}\rfloor}\sum_{q'=0}^{\lfloor\frac{n'}{2}\rfloor}\frac{m!n!m'!n'!}{j!q!j'!q'!}
 \frac{(-{\rm i})^{N+N'}\, 2^{\frac{3}{2}(N+N')-3(j+j'+q+q')}}{(m-2j)!(n-2q)!(m'-2j')!(n'-2q')!} \nonumber\\
 &\hspace{7cm}\times ({\cal J}^\pm_\ell)_{{\cal N},j,q;{\cal N}',j',q'}^{{\rm HG}}(k) \,,
\end{align}
with
\begin{align}
 \!\!({\cal J}_\ell^+)_{{\cal N},j,q;{\cal N}',j',q'}^{{\rm HG}}(k)
 &={\rm e}^{{\rm i}\ell(\varphi_{m,n}+\varphi_{m',n'})}\bigl(\partial_{h_{\rm x}}\bigr)^{m+m'-2(j+j')}\,\bigl(\partial_{h_{\rm y}}\bigr)^{n+n'-2(q+q')} \nonumber\\
 &\quad\times\bigl(1+\ell\partial_{h_{\rm z}}\bigr)^{N+N'+1}\, {\rm e}^{-\frac{(w_0\vec{k}_\perp+\vec{h}_\perp)^2}{8}} \nonumber\\
 &\quad\times F_{2(N+N'+1-j-j'-q-q')}\!\bigl(\bigl|h_{\rm z}-{\rm z}_R(k\hat\kappa)+\ell\tfrac{(w_0\vec{k}_\perp+\vec{h}_\perp)^2}{8}\bigr|,0\bigr)\Big|_{\vec{h}=0} \,, \label{eq:IHGp}
\end{align}
and
\begin{align}
 ({\cal J}_\ell^-)_{{\cal N},j,q;{\cal N}',j',q'}^{{\rm HG}}(k)
 &= {\rm e}^{{\rm i} \ell(\varphi_{m,n}-\varphi_{m',n'})}\bigl[1+{\rm sign}\bigl(\ell(N-N')\bigr)\partial_{h_{\rm z}}\bigr]^{|N-N'|}  \bigl(\partial_{h_{\rm x}}\bigr)^{ 2\{\frac{m+m'}{2}\} }\nonumber \\
 &\quad\times\bigl(\partial_{h_{\rm y}}\bigr)^{2\{\frac{n+n'}{2}\}}\,\Bigl(\frac{1}{2}\partial_{c_{\rm x}}\Bigr)^{\lfloor\frac{m+m'}{2}\rfloor-j-j'}\,\Bigl(\frac{1}{2}\partial_{c_{\rm y}}\Bigr)^{\lfloor\frac{n+n'}{2}\rfloor-q-q'}\frac{1}{\sqrt{c_{\rm x}c_{\rm y}}} \nonumber\\
 &\quad\times F_{|N-N'|+2\{\frac{m+m'}{2}\}+2\{\frac{n+n'}{2}\}}\bigl(|h_{\rm z}-{\rm z}_R(k\hat\kappa)|,\Sigma_{i=1}^2\tfrac{(w_0k_i+h_i)^2}{8c_i}\bigr)\Big|_{\vec{c}=1,\,\vec{h}=0}\,. \label{eq:IHG0}
\end{align}
Here, we made use of the shorthand notations $\vec{k}_\perp=k_{\rm x}\vec{e}_{\rm x}+k_{\rm y}\vec{e}_{\rm y}$ and $(k\hat\kappa)=k_{\rm z}-\omega$ introduced in Ref.~\cite{Karbstein:2017jgh};
${\rm sign}(.)$ is the sign function. Besides, $\lfloor n\rfloor$ is the floor function which gives as output the largest integer less than or equal to $n$, and $2\{\frac{n}{2}\}:=n-2\lfloor\frac{n}{2}\rfloor$.

Conversely, for {\bf Laguerre-Gaussian beams} we have ${\cal N}=\{l,p\}$, $N=|l|+2p$, and obtain
\begin{equation}
 ({\cal I}^\pm_\ell)_{{\cal N};{\cal N}'}^{{\rm LG}}(k)
 =\frac{1}{4}\sum_{j=0}^p\sum_{j'=0}^{p'}\frac{(-\sqrt{2})^{|l|+|l'|}}{ j!j'!}\binom{p+|l|}{p-j}\binom{p'+|l'|}{p'-j'} \,
 ({\cal J}^\pm_\ell)_{{\cal N},j;{\cal N}',j'}^{{\rm LG}}(k)\,,
\end{equation}
with
\begin{align}
 ({\cal J}^+_\ell)_{{\cal N},j;{\cal N}',j'}^{{\rm LG}}(k)&={\rm e}^{{\rm i}\ell(\varphi_{l,p}+\varphi_{l',p'})}\,2^{j+j'}
  \bigl(1+\ell\partial_{h_{\rm z}}\bigr)^{N+N'+1}\bigl(\partial_{h_{\rm x}}^2+\partial_{h_{\rm y}}^2\bigr)^{j+j'} \nonumber\\
  &\quad\times\bigl({\rm i}\partial_{h_{\rm x}}+{\rm sign}(l\ell)\partial_{h_{\rm y}}\bigr)^{|l|}\bigl({\rm i}\partial_{h_{\rm x}}+{\rm sign}(l'\ell)\partial_{h_{\rm y}}\bigr)^{|l'|}\,{\rm e}^{-\frac{(w_0\vec{k}_\perp+\vec{h}_\perp)^2}{8}}\nonumber\\
 &\quad \times  F_{N+N'+|l|+|l'|+2(j+j'+1)}\bigl(\bigl|h_{\rm z}-{\rm z}_R(k\hat\kappa)+\ell\tfrac{(w_0\vec{k}_\perp+\vec{h}_\perp)^2}{8}\bigr|,0\bigr)\Big|_{\vec{h}=0}
\end{align}
and
\begin{align}
  ({\cal J}_\ell^-)_{{\cal N},j;{\cal N}',j'}^{{\rm LG}}(k)&={\rm e}^{{\rm i} \ell(\varphi_{l,p}-\varphi_{l',p'})} \bigl[1+{\rm sign}\bigl(\ell(N-N')\bigr)\partial_{h_{\rm z}}\bigr]^{|N-N'|}\, \partial_c^{j+j'} \nonumber\\ 
 &\quad\times 
 \bigl({\rm i}\partial_{h_{\rm x}}+{\rm sign}(l\ell)\partial_{h_{\rm y}}\bigr)^{|l|}\bigl({\rm i}\partial_{h_{\rm x}}-{\rm sign}(l'\ell)\partial_{h_{\rm y}}\bigr)^{|l'|} \nonumber\\  
 &\quad\times \frac{1}{c}F_{|N-N'|+|l|+|l'|}\bigl(|h_{\rm z}-{\rm z}_R(k\hat\kappa)|,\tfrac{(w_0\vec{k}_\perp+\vec{h}_\perp)^2}{8c}\bigr) \Big|_{c=1,\,\vec{h}=0}\,.
\end{align}

\section{Conclusions and Outlook}

In this proceedings, we have determined analytic expressions for the one-loop photon polarization tensor in circularly polarized Hermite- and Laguerre-Gaussian beams, thereby complementing our recent study~\cite{Karbstein:2017jgh} for linearly polarized beams.
Our results are based on a LCFA of the Heisenberg-Euler effective Lagrangian for QED in constant fields, and thus are manifestly limited to slowly varying electromagnetic fields.
More specifically, our results neglect contributions of the following type, 
\begin{equation}
\sim\upsilon^2\Bigl[{\cal O}\bigl(\tfrac{2}{w_0\Omega}\bigr)+{\cal O}\bigl(\tfrac{1}{\tau\Omega}\bigr)+{\cal O}\bigl((\tfrac{\upsilon}{m_e})^2\bigr)+{\cal O}\bigl((\tfrac{\alpha}{\pi})^2\bigr)\Bigr]\,,
\end{equation}
where $\{\omega,|\vec{k}|,\omega',|\vec{k}'|,\Omega\}\lesssim\upsilon\ll m_e$; see Ref.~\cite{Karbstein:2017jgh} for the details.
The first term in the squared brackets arises from the restriction to the (leading order) paraxial approximation, the second one is due to the finite pulse duration, the third one refers to contributions beyond the LCFA, and the last one to contributions from higher loops.

As any paraxial beam can be decomposed into either Hermite- or Laguerre-Gaussian modes, our results are expected to be relevant for the study of optical signatures of QED vacuum nonlinearity in realistically modeled, circularly polarized high-intensity lasers pulses.

\ack 

F.K. is grateful to Matt Zepf for stimulating discussions.
The work is supported by Russian Science Foundation grant No. 17-72-20013.
%The research of E.A.M. is funded by the Tomsk Polytechnic University Competitiveness Enhancement Program grant, CEP-PTI-72/2017.

\end{document}